# Does the brain function as a quantum phase computer using phase ternary computation?

Andrew S Johnson[1] and William Winlow[1,2]

[1]Dipartimento di Biologia, Università degli Studi di Napoli, Federico II, Via Cintia 26, 80126 Napoli, Italia;
[2]Institute of Ageing and Chronic Diseases, The Apex Building, West Derby Street, University of Liverpool, Liverpool, L7 8TX, UK

**Abstract**

Here we provide evidence that the fundamental basis of nervous communication is derived from a pressure pulse/soliton capable of computation with sufficient temporal precision to overcome any processing errors. Signalling and computing within the nervous system are complex and different phenomena. Action potentials are plastic and this makes the action potential peak an inappropriate fixed point for neural computation, but the action potential threshold is suitable for this purpose. Furthermore, neural models timed by spiking neurons operate below the rate necessary to overcome processing error. Using retinal processing as our example, we demonstrate that the contemporary theory of nerve conduction based on cable theory is inappropriate to account for the short computational time necessary for the full functioning of the retina and by implication the rest of the brain. Moreover, cable theory cannot be instrumental in the propagation of the action potential because at the activation-threshold there is insufficient charge at the activation site for successive ion channels to be electrostatically opened. Deconstruction of the brain neural network suggests that it is a member of a group of Quantum phase computers of which the Turing machine is the simplest: the brain is another based upon phase ternary computation. However, attempts to use Turing based mechanisms cannot resolve the coding of the retina or the computation of intelligence, as the technology of Turing based computers is fundamentally different. We demonstrate that that coding in the brain neural network is quantum based, where the quanta have a temporal variable and a phase-base variable enabling phase ternary computation as previously demonstrated in the retina.

1) **Introduction**.

Traditionally a nerve impulse has been considered to be an electrochemical phenomenon with experiments dating back 250 years to Galvani **(**Bresadola 1998). However, this assumption has prevented contemporary consideration of the issues surrounding computation and assumes the temporal and communicative aspects of nervous activity can be resolved by electrical theory. In other words, much has been done to understand the biophysical mechanisms underlying action potentials, but not enough is known about neural computation, which is a separate issue.

The assumption that ionically based electrical communication within neurons is the fundamental processor of computation has inevitably led to models of both intelligence and computation being created using this technology in computing sciences and more recently in artificial intelligence (AI). Thus, contemporary models of nerve conduction rely on the original work of Hodgkin and Huxley (HH) and their excellent work on the action potential (Hodgkin and Huxley, 1952) that has led to the peak of the action potential as the temporal marker for computation and propagation of the action potential assumed by cable theory. The orthodox action potential (Fig.1) is comprised of a spike with a peak at about 0.2ms from its inception. It is evident from this curve that activation begins close to the resting potential. At resting potential, the sodium channel activates with little delay (Almog 2018) The scale below Fig 1 shows approximate distances along the axon indicating that no charge from the spike could affect activation as the main charge is prior to the point of initiation. In addition, the exponential rise of the Na+ current demonstrates that activation of the exponential release of ions commences with little or no charge. i.e. at threshold.  The molecular distances between ion channels far exceed the distance required to allow the level of charge (Hodgkin and Huxley, 1952) needed to activate progressive ion channels to achieve propagation.  This is in agreement of our earlier study (Johnson and Winlow, 2018) where distances taken from patch clamp studies confirm that HH cable theory cannot account for propagation. At the time of HH inter-channel distances were unknown.

The HH equations are a set of nonlinear differential equations that approximate the electrical characteristics of excitable cells and can describe the electrical potential caused by exponential passage of ions notably Na+ and K+ when Na+ enters through ion channels in the surface membrane. Later work indicates that some action potentials are also calcium dependent (Hayer 1981). The equations describe the itemised potential changes of these ions over time. Propagation of the action potential along the membrane is assumed in HH to be directly due to charge from the ensuing spike opening proximal ion channels. Opening and closing of the ion channels must result in morphological changes to the ion channels proteins entailing force on the membrane. We have previously suggested that this model is insufficient to explain the activation of channels (Johnson and Winlow 2018) and that the activation actually moves ahead of the charge at a position where the charge is ineffective. In the HH model, propagation is assumed to be a result of capacitance change creating enough charge to affect the next ion channel on the membrane. We dispute that this is possible (Johnson and Winlow 2018a, Johnson and Winlow 2018b) and have proposed an alternative theory for propagation (Johnson and Winlow 2018a).

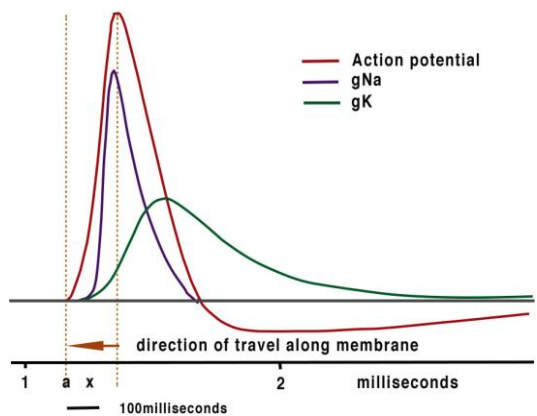

**Figure 1: The conventional textbook action potential showing the sodium and potassium currents with which it is associated.**

2) Action potentials play more than one role in CNS function

We do not dispute that HH action potentials are driven by the entry and exit of ions acting down their concentration gradients, as shown in Figure 1, or that action potentials serve a number of functions such as:
- wiring the nervous, sensory effector and neurosecretory systems during development (Pineda and Ribera, 2007)
- formation and maintenance of synaptic connections in the adult (Forehand, 2009; Dickins and Salinas, 2013; Andreae and Burrone, 2014)
- modulation of synaptic function during and after learning (Kennedy, 2016; Rama, et al, 2018).

However, it is not yet broadly accepted that action potentials are associated with underlying pressure pulses, known as solitons, (Johnson and Winlow, 2018). Taken together they form the action potential pulse (APPulse) which allows very rapid computation, as we have suggested for the retina (Johnson and Winlow, 2019). This is an important concept, given the plasticity and multiple formats of action potentials.

3) Timing Plasticity and error

The non-uniformity of both neurons and their transmission properties is an important determinant in the type of information conveyed by them and the possible types of computation available. The brain is a large mass of neurons whose coupling, connections and form are inherently plastic. This plasticity takes various forms depending upon the timing periods sampled. Neurons may change or be replaced over weeks and months, synaptic connections may change over minutes, and conduction across synapses changes after milliseconds, ion channels within the neurons may disperse over the membrane and are regularly replaced. Many forms of plasticity affect both the temporal position of spiking neurons and their amplitude, repetitive activation inevitably changes the concentrations across the membrane which go to define the shape and timing of the conventional spike (Figure 1). Any change to the temporal position of the spike peak will thus affect computation (Figs 2 and 4). Therefore, it is absolutely essential that for any computation to take place within a neural network this temporal plasticity must be negligible in comparison to the temporal timing. This is especially true when considering parallel processing within a network where parallel threads of information must be synchronised. For any useful consecutive computation to occur the structure must be stable, *within the relatively short timeframe of computation*, i.e. microseconds rather than milliseconds

**Action potential plasticity -** Action potentials are thought of as the means by which cellular communication takes place within the nervous system and serve to trigger secretions from nerve terminals. They are generated by powerful ionic driving forces created by metabolic pumps such as the sodium-potassium pump, which instigate the membrane potential (Fig 1). However, action potentials are plastic phenomena (Fig2.1), whose properties vary substantially from one neuron to the next (Winlow et al, 1982; Bean, 2007)(Fig. 2.2) and are often compartmentalised within neurons (Fig 3) (Haydon and Winlow, 1982) such that the action potentials of cell bodies, dendrites, axons and nerve terminals may be quite different from one another in terms of their ionic makeup. They should be considered as a signalling mechanism for the release of secretory products at a distance from the soma (Winlow, 1989). The excellent work of Hodgkin and Huxley (Hodgkin and Huxley 1952) (HH) in determining the ionic nature of action potentials has largely obscured the accruing evidence that the plasticity of action potentials in cell bodies, nerve terminals (Bourque, 1990; Spanswick and Logan, 1990; Winlow, 1985) and axons (Rama, et al, 2018) makes them unsuitable for computation within the nervous system (Winlow and Johnson, 2020). Indeed, action potential trajectories differ so much from one another that they have been used to classify different neuronal types in the neuronal somata of the pulmonate mollusc *Lymnaea stagnalis* (Winlow et al, 1982) and in vertebrates (Bean, 2007). In particular the variable position action potential peak is well documented (Bourque, 1990; Bean 2007) (Fig 2.1) and the maximum rates of depolarization ($\dot{V}_d$) and repolarization ($\dot{V}_r$) are highly variable phenomena and are clearly frequency dependent (Fig 4.1) as can be demonstrated using the phase plane technique (Holden and Winlow, 1982), which is very useful for determining the threshold of action potentials (Bean, 2007; Trombin et al, 2011; Li et al, 2014; Xiao et al 2018; Winlow and Johnson, 2020). Frequency changes result in a shift of the action potential peak and both $\dot{V}_d$ and $\dot{V}_r$ are modifiable by excitatory and inhibitory synaptic inputs (Winlow, 1985; Bourque, 1990; Bean 2007) (Fig 4.2). In addition, neurons lie close to one another in nerve trunks and central nervous systems, so that modification of the extracellular medium by neuronal activity may alter ionic concentrations, thus modulating action potential trajectories.

Binary computational models of nervous systems usually use the peak of the spike to initiate activity (Taherkhani 2020), but given the variability of V̇d, this is clearly an inaccurate method of computation. We have shown elsewhere that ternary phase computation is much more appropriate in modelling nervous activity where threshold is the instigator of the computational action potential (CAP): the three phases are thus: resting potential, threshold and the time-dependent refractory period, which is an analogue variable (Johnson and Winlow, 2017, 2018a,b).

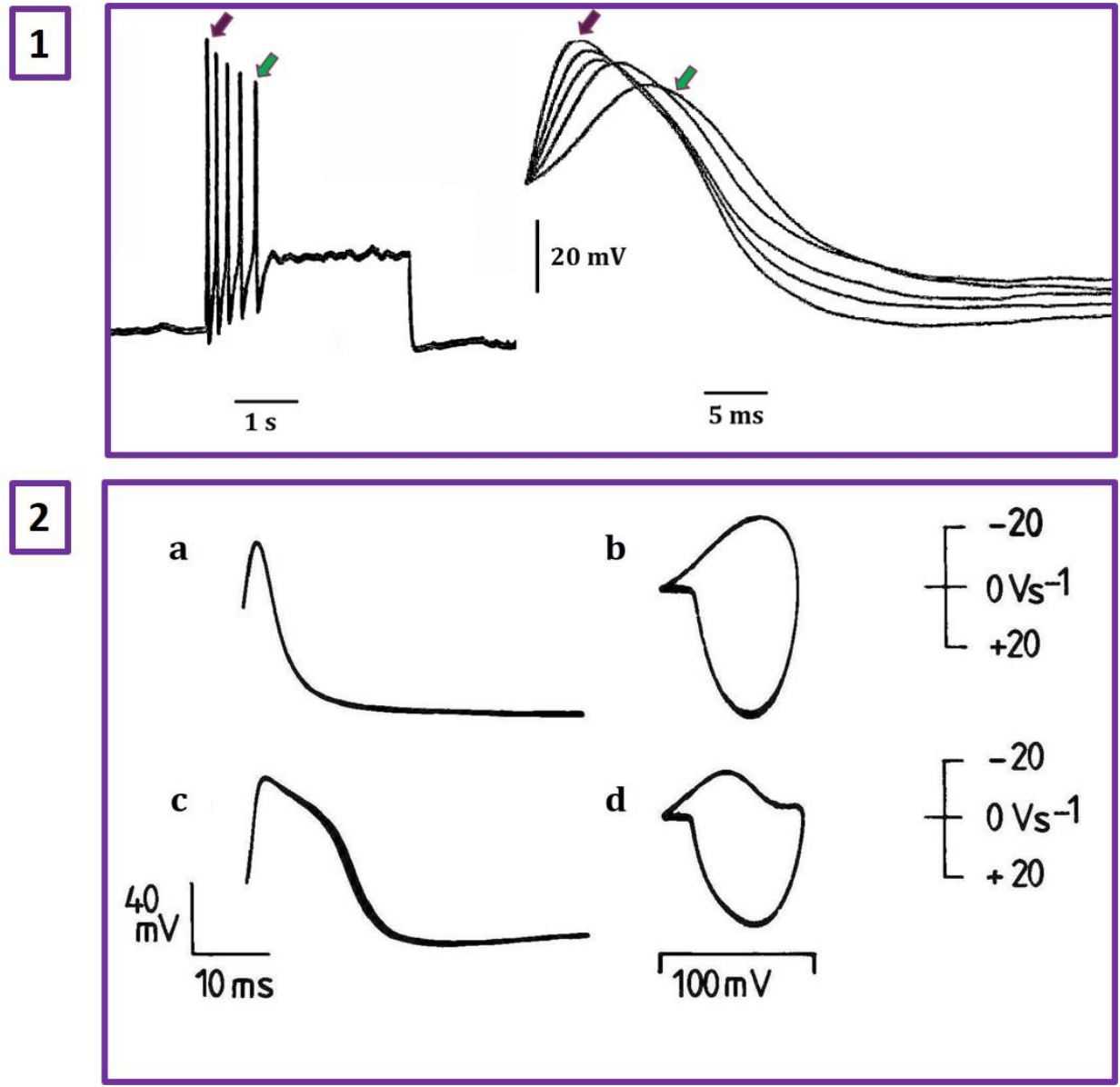

**Figure 2: Examples of action potential plasticity from identified neuronal somata of the great pond snail *Lymnaea stagnalis* (L.) (for locations and properties and functions (where known) of neurons and cell clusters, see Slade et al, 1981 and Winlow and Polese, 2014).**
**1)** Action potentials from a fast-adapting pedal I cluster cell, which was normally silent and was activated by a 3 s, 0.4 nA pulse injected into the cell via a bridge balanced recording electrode. The same five spikes are shown in each case: a) on a slow time base, b) on a faster time base. The red and green arrows indicate the first and last spikes respectively in each case. Note the temporal variability between spike peaks (previously unpublished data provided from William Winlow's data bank).

**2)** Different types of action potentials have different spike shapes (a & c) and trajectories as demonstrated in phase plane portraits (b & d). a&b) demonstrate the same type 1 action potentials from an RPeF cluster neuron, while c & d) demonstrate type 2 action potentials from an RPeB cluster neuron (adapted from Winlow et al, 1982 with permission). In the phase plane portraits, the rate of change of voltage (dV/dt) is plotted against voltage itself and the inward depolarizing phase is displayed downward, maintaining the voltage clamp convention. The technique is very useful for determining action potential thresholds (see Holden and Winlow (1982) for details of the phase plane technique as shown here).

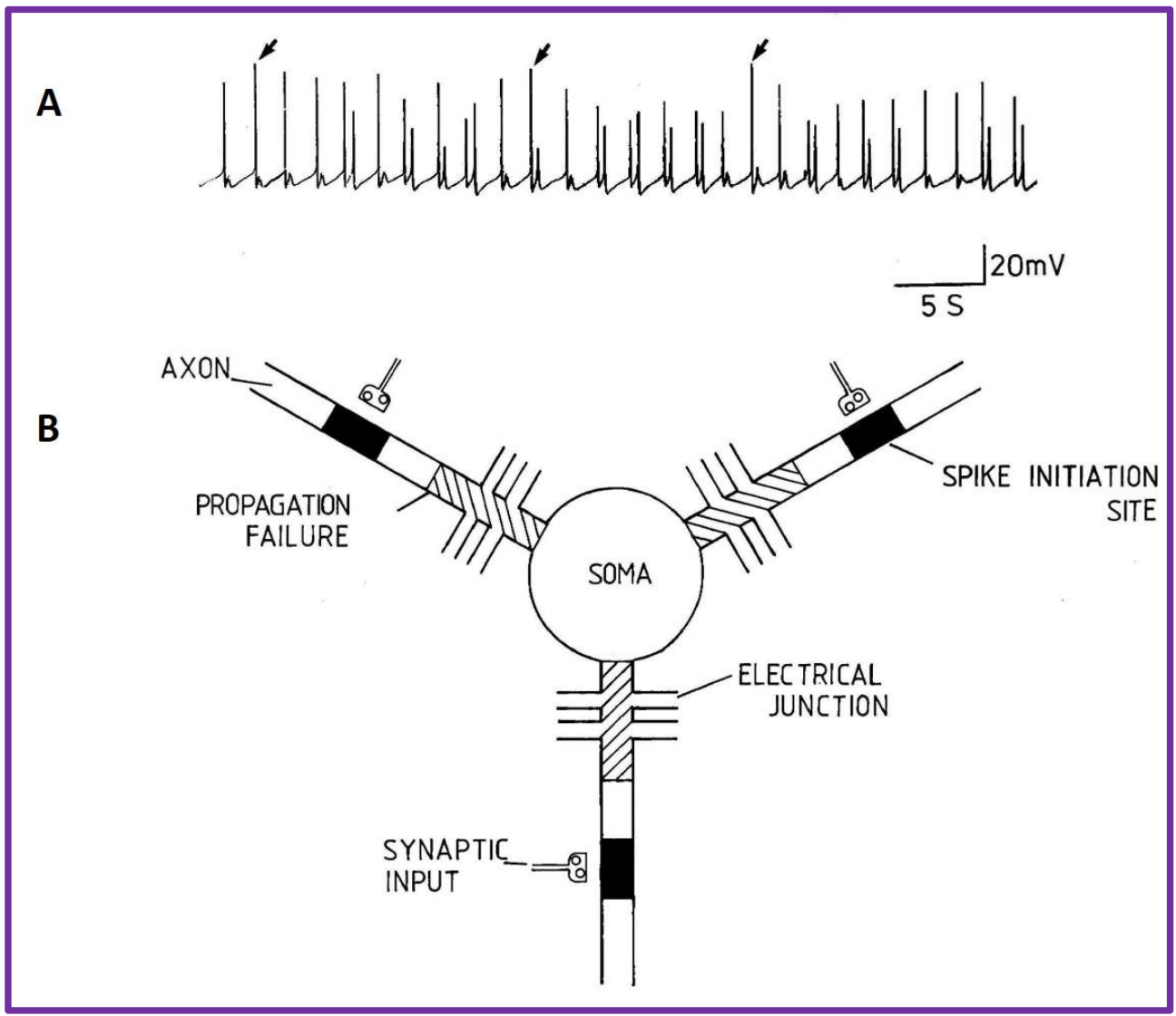

Figure 3: An example of compartmentalization of axons from a multipolar pleural D group neuron of *Lymnaea stagnalis*.
**A)** Spontaneous discharges recorded from the soma of a multipolar pleural D group neuron. The discharge varies between subthreshold depolarisations of varying amplitude to large overshooting action potentials (From Winlow, 1989, with permission).
**B)** Diagrammatic model of pleural D group neurons. The filled boxes in the axons represent spike initiation sites and the cross hatching represents regions of propagation failure. Each radial axon is spontaneously active and can act to generate spikes independently of the others. The electrical synapses appear to keep the axons functionally compartmentalised from one another (Reproduced, with permission, from Haydon and Winlow, 1982).

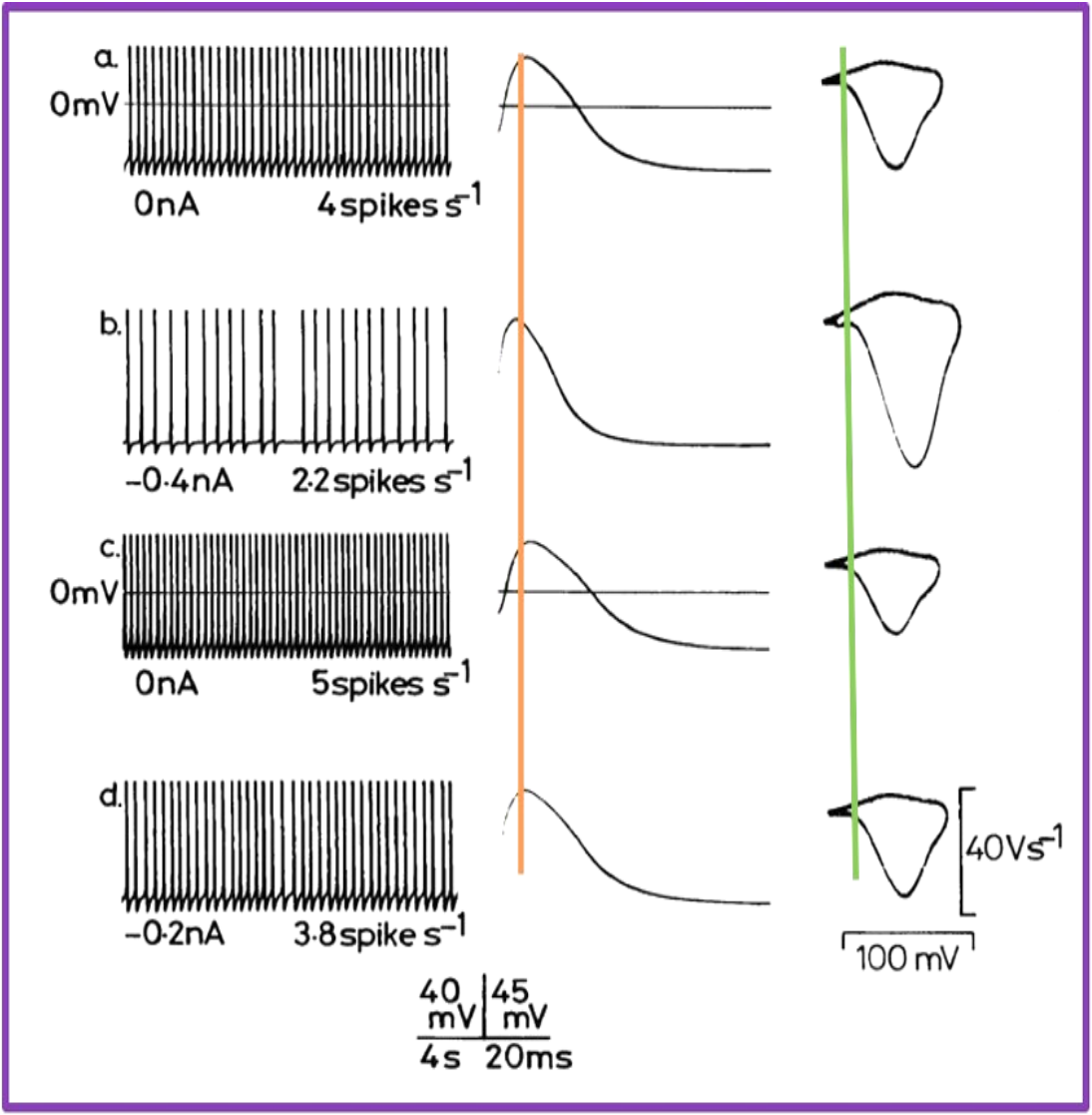

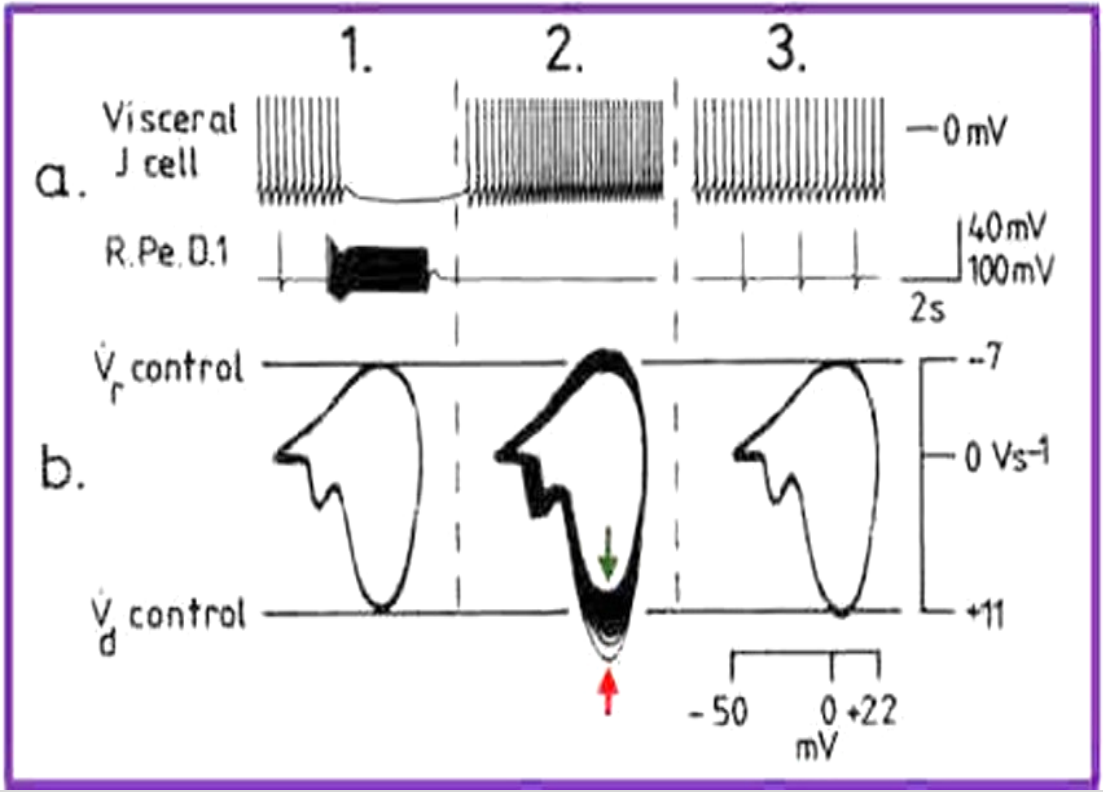

**Figure 4: Modulation of action potential frequency, shape and trajectories from identified *Lymnaea* neurons by (1) depolarising currents and (2) inhibitory synaptic inputs.**
**1)** In a right pedal A (RPeA) cluster neuron, action potential frequency, shape and phase plane trajectory are modified by maintained depolarizing and hyperpolarizing currents, which modify the action potential properties. All spikes in the traces at left are superimposed in the middle trace and in the phase plane portraits at right. Note how the action potential peak at $\dot{V}_d$ shifts either side of the superimposed orange line, while threshold (superimposed green line) remains temporally constant and $\dot{V}_d$ is even more clearly variable in the phase plane representations. Similar effects are produced by excitatory and inhibitory synaptic inputs (adapted from Winlow et al, 1982 with permission).
**(2)** Effects of monosynaptic i.p.s.p.s from the giant dopamine-containing neuron, RPeD1 (right pedal dorsal 1), on visceral J cell action potentials (for detail see Winlow and Benjamin, 1977). a) upper trace, J cell; lower trace RPeD1 (ac coupled and filtered). b) phase plane portraits of J cell action potentials: 1. pre-control; 2 phase plane portraits of 32 successive action potentials (peak of spike 1 denoted by red arrow, spike 3 by green arrow); 3) post-control of the last 10 action potentials shown above. After inhibition by RPeD1 both $\dot{V}_d$ and $\dot{V}_r$ were increased, but as the cell accelerated following inhibition both $\dot{V}_d$ and $\dot{V}_r$ declined below control values (modified from Winlow, 1985).

4) **Sequential (ST) and Parallel Network Computation (PNC).**

***Turing Machines*** - Almost all contemporary computers are designed around a Turing machine (Turing 1937), a mathematical model of computation that defines an abstract machine, which manipulates symbols on a strip of tape according to a table of rules. The programme is provided on successive sections of the tape each synchronised externally by a clock precise to each command. Whether in using an abacus to count numbers, or a modern computer to type a scientific paper the basis of computation remains identical. At its most simple any defined set of independent inputs leading to a defined set of outputs is computation. Figure 5a illustrates clock timed Turing machines in binary (0,1) and in ternary notation (-1, 0, +1) (Figure 5c).

The philosophy of Turing compatible machines when conceived in 1936 (Copeland 2004) was heavily influenced by not only the hardware architecture available, but also the applications to which the technology could be applied. Research into computation has always followed man-made hardware and the applications it can deliver. The establishment and acceptance of the action potential as the mechanism for nerve transmission was unavailable to Turing and his contemporaries. Unfortunately, contemporary research has largely ignored the fundamental differences between a Turing compatible machine and the brain. Recent attempts to re-imagine the brain as a Turing compatible environment are therefore a product of this process. Each step of a Turing compatible computer programme is timed by repetition of this logic by clock-steps. In parallel processing each parallel thread must also be precisely synchronised to produce logical output. Furthermore, conventional computation in neural networks, both real and artificial, relies upon selective gating of distinct routes through the network that are timed. For consistent computation to occur among parallel inputs successive parallel inputs must be synchronised precisely so that their activities can be executed.

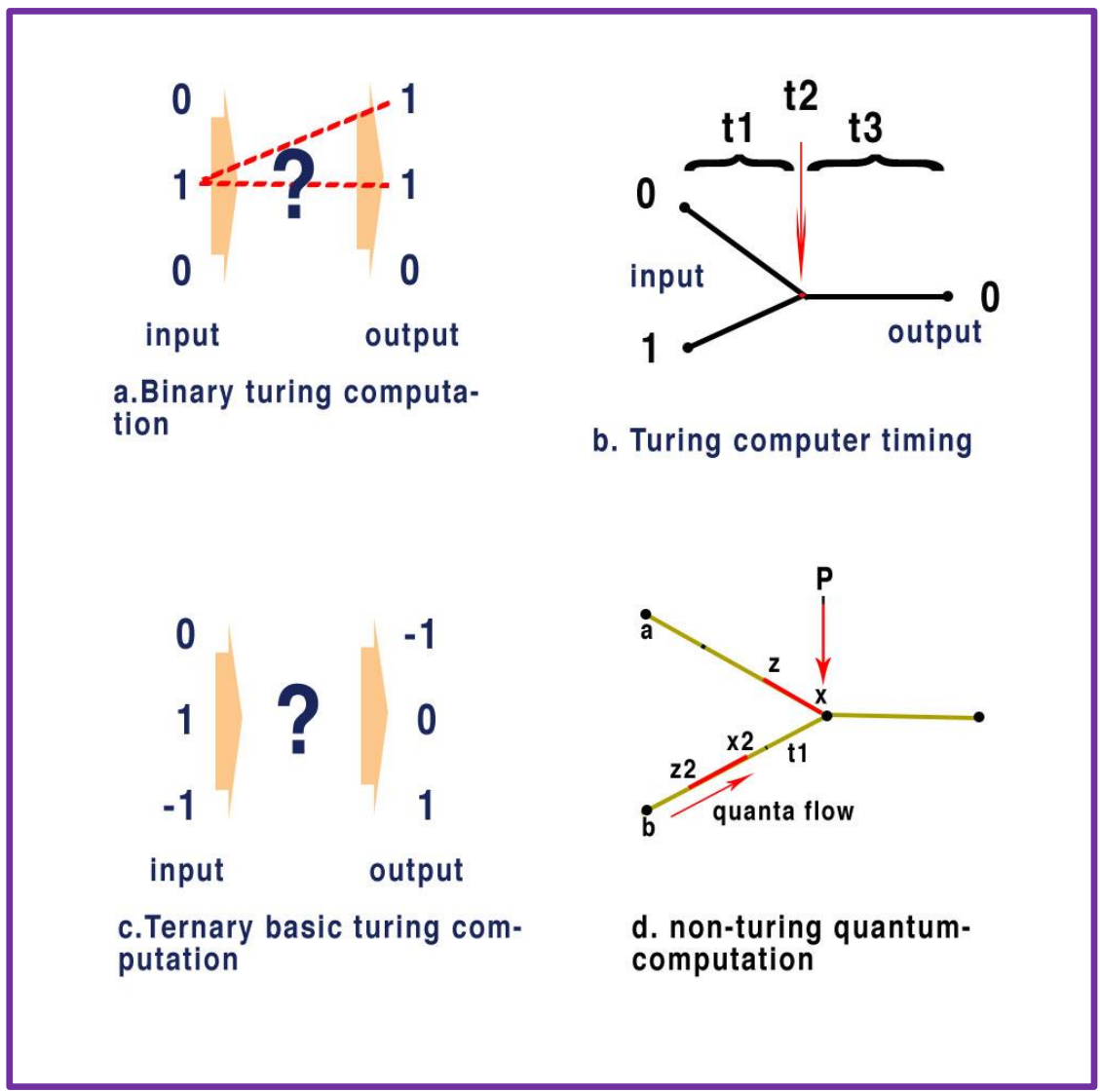

**Figure 5: In quantum computation network timing does not require an external clock.** This form of computation only requires input to reflect consistent outputs for given values and here we compare it with Turing computation where clock speed synchronises input with output. **(a)** is an illustration of conventional binary computation of inputs and outputs. The exact network that separates the input from the output is indicated by (?) and the most obvious pathways are shown as dotted lines and gating within the network, (?) forms this pathway. This gating represents a single command and must be executed within the set time allocated between the input and output as defined by some external clock. (?) may represent a complex network, flow through the network is defined node to node by precise timing between those nodes in exactly the same way and the outputs are always synchronised by clock-speed (as in the central processing unit of a computer). b) illustrates binary computation in a conventional space split between input nodes with binary properties of (0) and (1) respectively and giving a result of (0). (t1) represents the timing from input nodes (0) and (1) respectively. (t2) represents the time taken for the gated node at (t2) to react to the inputs (0) and (1) according to a set program that defines the output from these specific inputs as being (0). (t3) represents the time from the gated output to the exit node. Quantum computing applied to 5 (b) combines t1 + t2 to represent one quanta of binary information in a parallel network. In a conventional neural network computer as used in AI or SpiNNaker (Furber et al 2007) an emulator using spiking AP (Aboozar et al 2020), where latencies of timing are fixed and arranged so that (t1) and (t3) are uniform t1 + t2 +t3 are equivalent to clock speed effectively synchronising each command. This is the mechanism apparent from the definition

of a Turing type machine and evident in all practical applications of modern computing. **(c)** is an illustration of a ternary Turing machine, note synchronisations are performed by external clock and this is not applicable to a brain neural network. **(d)** illustrates a non-Turing quantum machine such as evidenced in the brain neural network. P is a point of convergence thus computation in the network and timing between nodes a and b to P of quanta (in red) is determined by the speed at which quanta flow between the nodes (black dots). Computation at P depends upon the quantal information (which may be any base), the mechanism at P, and the timing of interference between quanta. In a binary Turing machine **(b)** the quanta are fixed between t1+t2 by clock speed. In a brain neural network timing is defined by the speed of neurons and not by a clock. For the APPulse or action potential, x to z, and x2 to z2 are represented by the threshold to refractory periods of each quantum respectively: computation at the P is defined by the timing of each quantum and how they react on collision. In the case of the action potential and the APPulse collision between threshold and refractory periods results in annulment of the succeeding quanta. This results in computation through the network. This is a fundamental process in quantum computation and can be applied to all base and temporal computation in a neural network where the rules at P may differ.

### 5) Quantum Phase Computing - Another type of synchronisation for the network

Quantum phase computing occurs when temporal phase quanta, comprising base information, interact to provide a consistent output. This defines theoretical computation but does not describe the physical components needed. All modern-day computers can be visualised as quantum phase computers where the base is binary and the clock timing defines the temporal position of the quanta. In contemporary quantum computing this takes place at the subatomic level at very low temperatures and is very fast. Nevertheless, the basic computational theory is identical to that of the nervous system as we describe below. In the nervous system quanta are in the form of quantum ternary structures (i.e. the CAP is a temporal quantum of ternary information). The CAP in the form of either the action potential or the APPulse, are quantum phase ternary structures able to interfere and synchronise within a parallel network of neurons. Synchronisation can be achieved by quantum interference in a parallel network with almost any base coding of the information. In a conventional Turing computer synchronisation occurs in binary coding because clock speed is always equal to the time taken for each phase. The levels of synchronisation are a function of the temporal precision of impulses as they collide and their format.

Quantum phase ternary computing, as in the proposed retinal model (Johnson and Winlow 2019 and see below) is just one of an infinite number of quantum phases capable of computing depending upon phase length and base. Quanta in each case will have the form: t (base). Time t is a temporal phase variable and base can be any base. (t) is a time constant at the point of computation equivalent to relative clock speed in a Turing machine. In a Turing machine the time (t), the clock time, is the same as the phase in each computation so (t) is 1, if base is base 2. However, this is not the case in the nervous system where there is no centralised clock, because the effective clock speed (t) varies between points of convergence in a neural network as refractory periods will change at each convergence. Connected nodes in a brain neural network are therefore computing in different clock frequencies. Furthermore, in parallel processing (t) must be a consistent for each node so that during computation plasticity does not affect the network environment, as illustrated in Figure 5d.

**Timing vs Synchronisation -** Figure 5a illustrates some of the rules of parallel computation that occur within a network. The hidden neural network (?) is a combination of nodes that form into a parallel network. Parallel computation has substantial advantages over consecutive computation in terms of speed and the ability to synchronise. The values of changing parallel inputs must always reflect the synchronised changed output. The precision of synchronisation is fundamental to effective computation. For an efficient parallel network any input combination must be capable of creating a unique output and the process must be replicable. For computation to occur there must be interference between the distinct sets of inputs as they pass the nodes on the network. A pathway must be available from each input that reflects the collisions and programming within the network. In AI, a network can have programmed rules of pathways according to clock-timing. However, in the brain, clock-timing is unavailable and so another rule must exist to synchronise activity.

Figure 5a deliberately does not specify the processes or timing that must take place as indicated by the ? symbol. This implies that more than one process may take place and that further divisions of time may be present within the system. A large neural network such as the brain will have billions of connections, but to reduce error and maintain efficiency the number of components must be kept to a minimum. Synchronicity to enable successive commands in the case of a Turing machine is by a clock. This is possible because time between processes in a Turing system is not phase dependent but absolute time dependent and thus determined by clock speed. Each process in a conventional computer is therefore separated equally by time. The rule that synchronicity of processing must be centrally timed cannot apply within the central nervous system of an animal where peripheral ganglia, such as the retinal ganglia, have the ability to compute independently but must synchronise with the central nervous system. Thus, the combined output from all the neurons in the optic nerves must synchronise for us to understand the whole picture.

### 6) Neurons in Biological Neural Networks

In contrast with an artificial neural network, a real neural network (RNN) is comprised of many neurons whose function follows their form and where neuronal morphology and function are interrelated and depend on each other (Ofer et al 2017, Grbatinić et al 2019). Detailed analysis of the membrane structure and function have been discussed elsewhere (Hodgkin and Huxley 1952, Johnson and Winlow 2018a, Johnson and Winlow 2018b, Johnson and Winlow 2019). The transmission of information takes place along the membrane of the neuron and there is a finite time taken for information to pass from one point to another, this is often termed latency. A typical speed of an action potential along an unmyelinated axon in the CNS is about 0.3 m/s. From this the minimum distance between the start of the action potential and the peak of the spike can be calculated, interferometry has recorded action potential at about 30mm/s (Boyle et al 2019, Ling et al 2018). Previously(Johnson and Winlow 2017) we discussed how the phase ternary action potential both synchronised and corrected for error. Later we identified that the action potential exists as a phase ternary pulse (Johnson and Winlow 2018a, Johnson and Winlow 2018b, Johnson and Winlow 2017) and is defined as such by HH (Hodgkin and Huxley 1952).

The action potential is a base 3, phase ternary structure (Johnson and Winlow 2018a, Johnson and Winlow 2018b, Johnson and Winlow 2017). The structure resembles that of a Qutrit (Almog et al 2019 Xiao 2013) with the exception that an action potential refractory period has no effect on the resting potential – both are similarly capable of computation. We have indicated that the action potential is always accompanied by a synchronised pressure pulse (Johnson and Winlow 2018a, Johnson and Winlow 2018b, Johnson and Winlow 2019) which we refer to as a soliton. Furthermore, deformations to the membrane in the form of a pressure pulse have recently been confirmed with interferometric imaging (Boyle et al 2019, Ling et al 2018). We consider that for propagation to occur it is probable that this pressure pulse does not need to form a soliton only a disturbance in the membrane sufficient to open adjacent ion channels (Johnson and Winlow 2018b). This pressure pulse is sufficient to account for conduction of information in non-spiking, spiking neurons and in hyperpolarising cells such as the cones of the retina. The temporal precision of this synchronised pulse is much greater than that of the HH action potential as its speed is determined by the structure of the membrane that has a rate of change many times slower than that of either computation or even action potential conduction.

Basically, the soliton activates the ion channels that then add entropy to the pulse; the speed of the pulse is then defined by static membrane components. Temporal plasticity of membrane transmission occurs at a far slower rate than that of the ionic exchanges in HH. Temporal error is therefore minimised in the APPulse. The component structures of the APPulse were then deconstructed into computational component parts to form The Computational action potential CAP (Johnson and Winlow 2017 ). The CAP is a mathematical representation of a ternary quantum pulse where, during a collision of two impulses, if a threshold crosses a refractory period the threshold is annulled. The CAP is equally valid for ternary quantum computation by either HH or the APPulse. The difference between the two lies in the temporal precision of successive impulses up to 10,000 times greater with a pressure pulse than with HH and cable theory (Johnson and Winlow 2018). Importantly the CAP assumes that the temporal pulse starts on activation and not from the spike peak as shown in Figures 2.1 and 4.1.

**Activation of propagation -** As shown above, assuming the peak of the action potential to be the timing cue of the nerve impulse is a fundamental oversimplification of the mechanics of propagation because the underlying statistical event triggering the action potential must be the point of temporal precision for computation and synchronisation, i.e. threshold. To be effective across parallel inputs determined by phase, temporal precision is critical to phase computation. The accuracy of phase at the point of convergence must take place within a substantially reduced timeframe where temporal plasticity of membrane micro-structure is close to zero.

**Parallel Computation -** One functional test of whether the transmission of information is performed by HH cable theory or by the APPulse is whether they are capable of computation within the known neural networks of the body. Neural networks in the brain process information between parallel inputs across disordered networks (Fig. 6). The phase at which each CAP arrives at a point of computation is thus a quantum of ternary information. CAPs moving within a neural network compute by collision diffraction along specific pathways defined by temporal geometry forming patterns and changing outputs (Johnson and Winlow 2018a, Johnson and Winlow 2018b).

Computation of CAPs in a neural network occurs naturally through the phenomenon of phase cancellation at each node (Johnson and Winlow 2018), where CAP's interfere with each other. The mathematics of these interferences is defined by the precision of the activation point of the CAP, i.e. what is assumed to be a threshold in HH. To understand this crucial element of parallel processing it is necessary to look more closely at the molecular level of the components. The effects of interference between CAPs is illustrated in Figure 5. This illustration is equally valid for computational precision across multiple neurons in a network where interference must temporally synchronise. In figure 6(d) CAP's flow from left to right across the surface of the three-dimensional axon of the neuron. When the activation-threshold encounters a refractory section of membrane it is annulled.

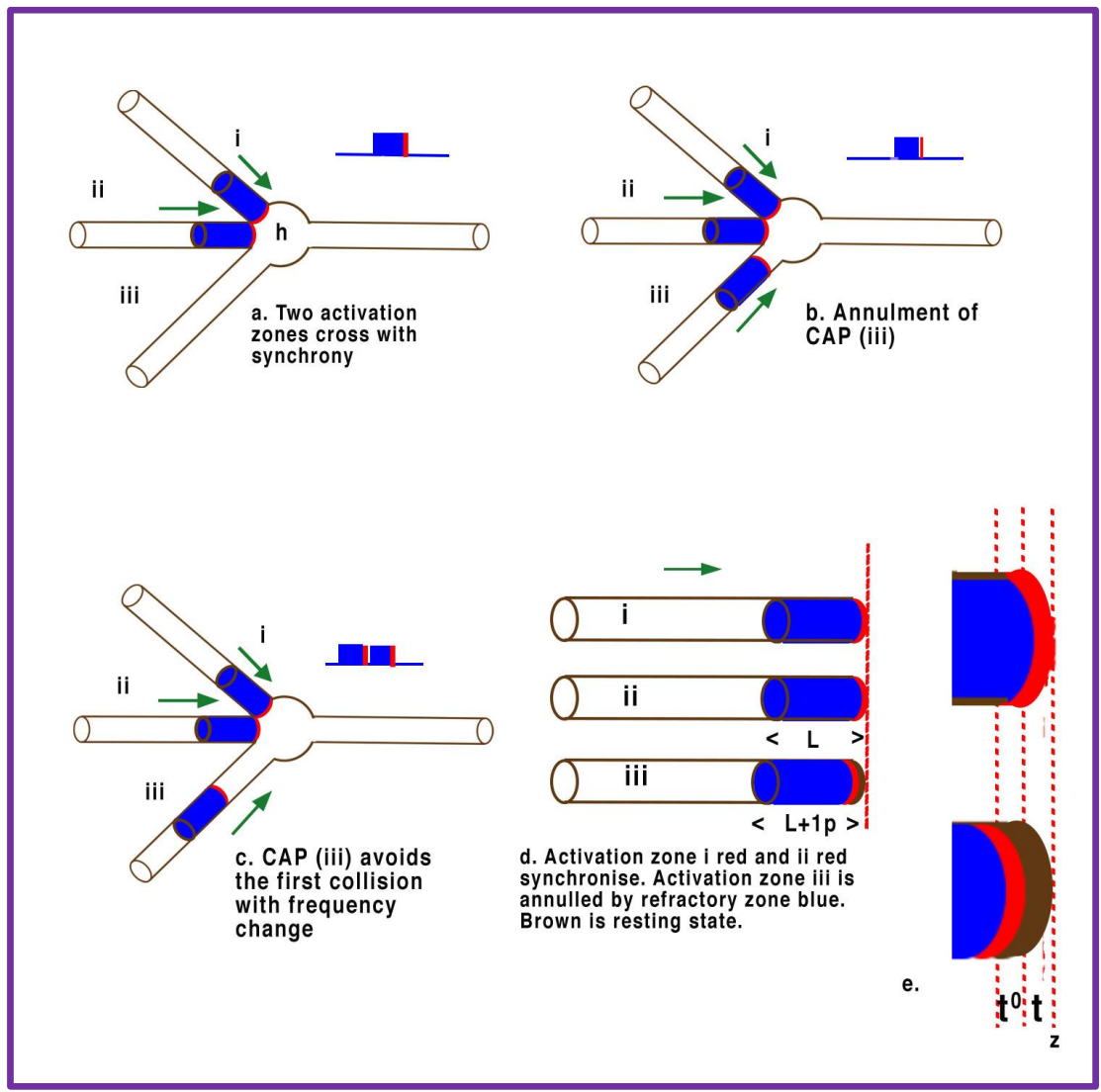

**Figure 6: Examples of quantum phase ternary interference between CAPs**. (a, b, c, d, e) are illustrations of CAP's travelling along the surfaces of the neuron membrane just before collisions at the axon hillock (h). The threshold of each CAP is highlighted in red while the refractory period is coloured blue. These are not to scale. (a) illustrates a single neuron with three axons converging on an axon hillock or cell body. Two CAP's (i) and (ii), move in the direction of the axon hillock where they will collide. In this case both CAP's (i) and (ii), are in phase with the thresholds overlapping. These CAPs will fuse and continue a single CAP.
(b) the same two CAP's (i) and (ii), are in phase with the thresholds overlapping and will fuse and continue as one CAP as in (a), but a third CAP (iii) arrives slightly after the other two and its threshold encounters the refractory period of the other two and is annulled.
(c) the same two CAP's (i) and (ii), are in phase with the thresholds overlapping. These CAP will fuse and continue as one CAP as before, but in this case the threshold of the third CAP (iii) arrives at the axon hillock after the first two have fused and their respective refractory periods have no effect on it. The result is that two CAPs will pass into the axon at right. In (d) (L) represents the refractory time of each CAP, whilst in (e) (t) represents the timing of the activation-threshold at (z), the point of computation. In a parallel network ((d)) where CAP's converge any time greater than (t) and less than (L) will result in phase change. This phase change changes the distance between successive CAP's and therefore frequency. Where activation-thresholds become desynchronised phases become annulled.

7) **The Retinal model of neural computation**

In a previous paper (Johnson and Winlow, 2019) we described the neural coding in the retina and further details and references are to be found in that paper. Figure 7 (b) is a schematic diagram of the relevant central elements of the retina. Light from the right of the diagram falls on grouped light receptors GLR (large coloured ovals that can be rods or cones). These light receptors are connected through a static array of bipolar neurons to retinal ganglion cells (RGC). Light receptors have an average of 12 per group and up to 25 that connect to a single bipolar cell whose output is usually observed in the RGCs (Behrens 2016). When all other connections are supressed each ganglion cell is activated by connected light receptors as shown in the diagram. Circular patterns of light receptors lie adjacent to one another or overlap each other and each is connected to a ganglion cell.

**Observations from the Retina -** Adjustments in light intensity results in a corresponding frequency change of action potential in the RGCs. The frequency of synaptic discharge from the cones is also similar to that of output t and t1 in Figure 6 (b). The frequency of discharges of all the light receptors attached to a bipolar cell result in a mean change to output frequency RGC (t) and (t1) respectively. Light on the retina and mean light receptor discharge frequency is therefore proportionally related to output at the RGC.s. In figure 6b) changes in light intensity on the grouped light receptors results in a mean frequency change of action potential at the output from the bipolar cells (t1). For this computation to occur there must be a point of computation (Johnson and Winlow 2019).

There appears to be only one mathematical mechanism that can generate the mean frequency of parallel streams of action potentials arriving at the LGN, i.e. mean sampling at the point of convergence of the bipolar cells. At the point of convergence, the refractory period of one CAP will block any succeeding CAPs for that period of time. This refractory period is dependent upon the membrane constituents and its timing is critical to computation. In a parallel system where CAPs converge when the first CAP passes through a point its refractory period blocks further action potential. When this CAP's refractory period ends there is a space in time before the next CAP's activation threshold, this is equivalent to the mean sampling of cone CAPs.

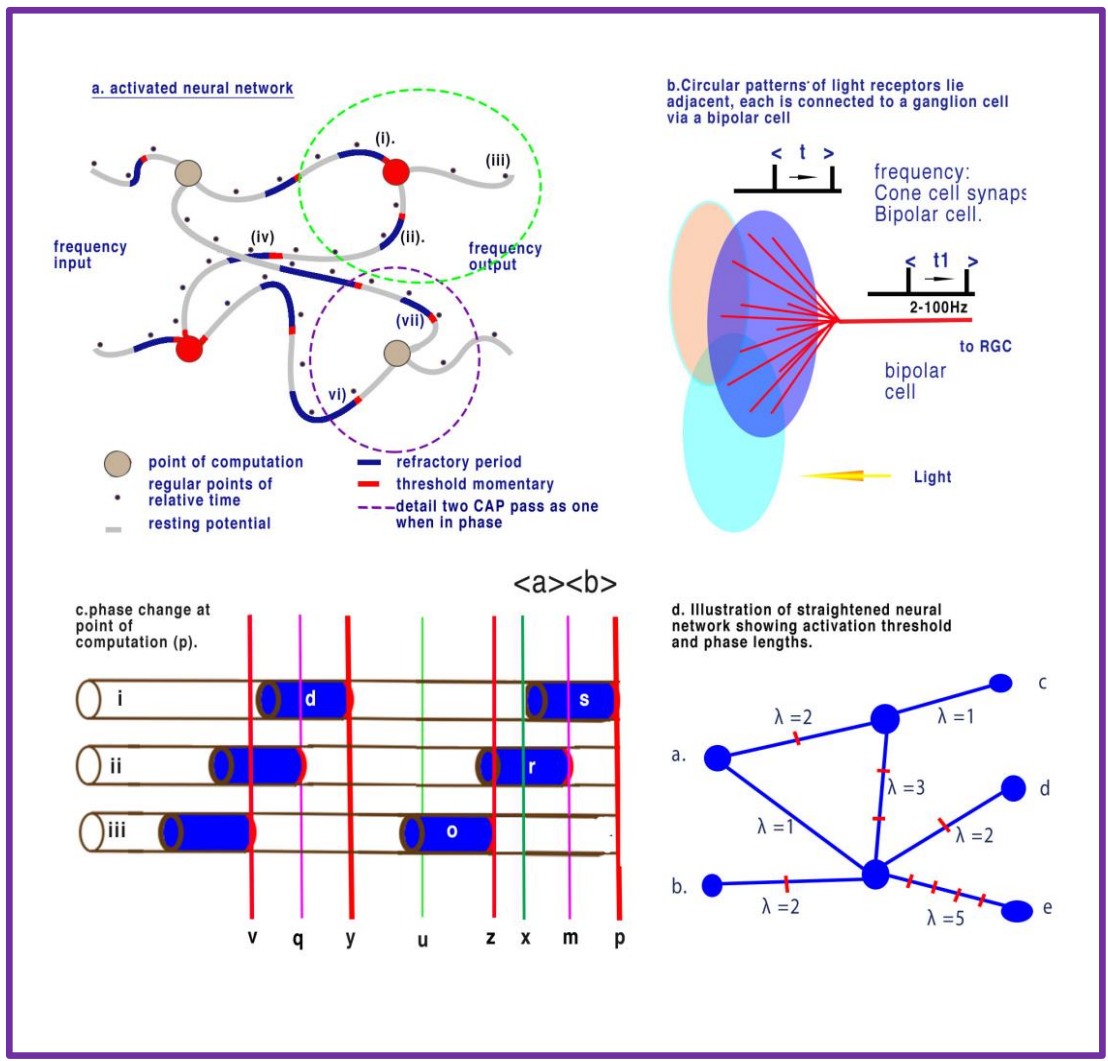

**Figure 7: Neural networks in the retina.** (a) Illustration of a brain neural network showing changes in latency between nodes (from Johnson and Winlow, 2019). (b) shows 12 cones (blue lines) converging upon a single bipolar cell (red lines). Here time, t1, changes relative to t. (c) is an enlargement showing a succession of CAPs arriving at the point of computation (p). During fixed light intensity each cone will cause CAP's to flow towards the point of computation (p) and CAPs are moving from left to right. Activation-thresholds that allow a CAP to continue after computation are shown by the red lines. All CAP in neurite (ii) are annulled. The time (p) to (x) is equivalent to the refractory period of the CAP of neuron (i). Refractory period (s) is selective and all other CAPs are annulled during this period. The result is that time between points of computation (z) to (p) is dependent upon only the time from the end of refractory period (x) to the next activation-threshold (z). This is then reflected in the spike timing from the RGCs. The interval (x) to (z) over any time period is a mean sample when calculated between successive CAPs. This results in the calculation of mean sampling by quantum phase ternary computation. Before collision the timing of each successive CAP along a neuron can be considered as a minute change in frequency. (d) illustrates a straightened disordered brain neural network such as in (a). Frequencies of inputs (a) and (b) result in changes to frequency outputs (c) (d)(e). In the network of (d) distances have been resolved into phase lengths of quanta. The network represents computation between different time dimensions illustrated by $\lambda$. As quanta arrive at different time at each node (between nodes quanta may arrive asynchronously but are digitised) convergence and interaction will take place where phases differ. This interaction will result in distinct changes in frequencies of output for distinct changes in frequencies of input.

**Digitisation of light receptor outputs by phase -** Using the known speed of CAP in the bipolar cells and the frequency of discharge, the minimum distance between each successive discharge can be calculated. The frequency rate of discharge from the bipolar cells has been measured at 2-100 Hz. Velocity of action potentials vary from neuron to neuron, measurements for an unmyelinated axon vary from over 25ms$^{-1}$ in the squid to 0.3 ms$^{-1}$ or below in brain tissue. The neural networks of the CNS and retina contain small unmyelinated neurites and the smaller figure is used in our calculations. The velocity of action potential for an unmyelinated axon has been measured by interferometry in nerve tissue at about 0.3ms$^{-1}$ (Boyle et al 2019).

**Calculation of precision -** The maximum frequency of the CAP is determined by the timing of the refractory period. If the maximum frequency is 100 Hz and the speed is 0.3m/s then the distance from the threshold-activation to the end of refractory period is: 0.3/100 = 0.003m or a time of 1ms. This figure corresponds to observed measurements of the absolute refractory period (Purves 2001). The activation mechanism for activation-threshold is likely to be timed less than 1-10 microseconds.

Single action potentials are not temporally accurate to less than 0.1 milliseconds measured from the spike peak, or even threshold, so computation in the eye is incompatible with the action potential, where reliability of temporal measurement is in order milliseconds. However, a pressure pulse or soliton is formed at the molecular level of the membrane and travels through the membrane at a constant speed with a greater accuracy than required (Heimburg and Jackson, 2005) for quantum phase ternary computation.

For the observed changes(Grimes 2014) in frequency when light shines on the cones, CAPs must be formed and compute at the convergences of the bipolar cells and the CAP must have an activation-threshold of below 10 microseconds. This timing is critical when we consider attempts to mirror nervous communication using models of spiking neurons obeying HH cable theory. In parallel processing the temporal precision is critical to operation. In a conventional view of nerves where computation takes place at the synapses

**Information contained between nodes -** Figure 5(b) illustrates a conventional binary neural network. During time, as defined by clock speed, one bit (0 or 1) is connected to the output. In a neural network this means that node to node contains one bit. Figure 7(c) illustrates quantum ternary phase pulses represented by CAP. If each refractory period is separated digitally into 12 (as above) then CAP (s) can be subdivided into 12 positions of phase change. In Figure 7(a), CAP (r) is annulled because its activation-threshold at point (m) will cross the refractory of CAP (s). The position of (m) is critical to computation because phases from (x) to (z) have been redacted. Each of the 12 subdivisions of the refractory period therefore code for one trit of information. In conventional computing terms the effective clock speed of computation for the bipolar cells is about 10 microseconds during which time each neuron connected to a convergence conveys 1 trit of information base 3. The space along a neuron of two impulses conveys base 9 information. The information contained is therefore much greater than possible with binary. There are many areas of the nervous system where connections similar to that of the bipolar cells are apparent for example the bipolar connections of the auditory system(Petitpré 2018).

8) **Quantum phase ternary computation in a network.**

Figure 7 d illustrates a straightened disordered brain neural network such as figure 6a. Frequencies of inputs (a) and (b) result in changes to frequency outputs (c) (d)(e) as described in Figure 7c. The nodes are the convergences. In Figure 7c distances between all nodes have been annotated with the phase length of the frequency $\lambda$ of CAP showing different distances. The phase length is the distance between two CAP activation thresholds (analogous to spike timing). Between any two nodes this phase length will remain constant during minimum plasticity, as it is dependent upon the integrity of the membrane. Between (a) and (c) $\lambda$ = 3 so there are 3 CAP. Between (a) and (e) $\lambda$ =6. Time for CAP between (a) and (c) is half that of (b) to (e). If the frequency of (a) changes from 1 to 2 the frequency of (a) to (c) changes 3 to 6 while frequency of (a) to (e) changes 6 to 12 or a change of 3 or 6 Hz depending upon the route. Interference between quanta in this system annul and change the direction of CAP within this network. As the frequency changes at inputs (a) or (b), output frequencies at (c) (d) and (e) change accordingly such that they provide a unique reference to the frequency inputs.

## 9) Conclusions

- In neural computation the action potential peak is unreliable for calculations because of action potential plasticity. Threshold is much more clearly temporally defined.

- The frequency of action potentials at the retinal ganglion is a result of interference between action potentials temporally measured only from the point of activation-threshold eliminating peak action potential-timed computation. In addition, the temporal accuracy of threshold activation must be less than 10 microseconds indicating that the Hodgkin Huxley action potential alone is incapable of computation.

- At the set point on the membrane where activation takes place charge from the spike is minimal. Thus, activation is the cause of the action potential and must be responsible for adjacent further activation.

- Computation during plasticity results in error in a neural network, which must be redacted, and we have proposed a method of phase ternary redaction. Spike timed computation is untenable because the spike arrives after the activation-threshold and the spike is a plastic phenomenon.

- The implication in terms of computer science is that in a parallel neural network, Turing based machines are a small subset of Quantum Phase Computing.

- Computation within brain neural networks is most likely by quantum phase ternary process, distinct from and much more precise than the action potential.